\documentclass[prb,preprint]{revtex4}
\usepackage[dvips]{graphicx}
\newcommand{\beq}{\begin{equation}}
\newcommand{\eeq}{\end{equation}}

\begin{document}

\title{Crossover between a Short-range and a Long-range Ising model }
\author{Taro Nakada$^{1,2}$, Per Arne Rikvold$^{3}$,
Takashi Mori$^{1,2}$, Masamichi Nishino$^{4}$, and Seiji Miyashita$^{1,2}$
}
\affiliation{
$^{1}${\it Department of Physics, Graduate School of Science,
The University of Tokyo, 7-3-1 Hongo, Bunkyo-Ku, Tokyo 113-8656, Japan}  \\
$^{2}${\it CREST, JST, 4-1-8 Honcho Kawaguchi, Saitama 332-0012, Japan}  \\
$^{3}${\it Department of Physics, Florida State University, Tallahassee, Florida, 32306-4350, USA}  \\
$^{4}${\it NIMS, Tsukuba, Ibaraki 305-0047, Japan}}

\begin{abstract}

Recently, it has been found that an effective long-range interaction is realized among local bistable 
variables (spins) in systems  where the elastic interaction causes ordering of the spins.
In such systems, generally we expect both long-range and short-range interactions to exist.
In the short-range Ising model, the correlation length diverges at the critical point.
In contrast, in the long-range interacting model the spin configuration is always uniform and the correlation length is zero.
As long as a system has non-zero long-range interactions, 
it shows criticality in the mean-field universality class, 
and the spin configuration is uniform beyond a certain scale.
Here we study the crossover from the pure short-range interacting model to the long-range interacting model. 
We investigate the infinite-range model (Husimi-Temperley model) as a prototype of this 
competition, and we study how the critical temperature changes 
as a function of the strength of the long-range interaction. 
This model can also be interpreted as an approximation 
for the Ising model on a small-world network.
We derive a formula for the critical temperature as a function of the strength of the
long-range interaction.
We also propose a scaling form for the spin correlation length
at the critical point, which is finite as long as the long-range interaction is included, 
though it diverges in the limit of the pure short-range model.
These properties are confirmed by extensive Monte Carlo simulations.
\end{abstract}

\date{\today}
\maketitle

\section{introduction}

Divergence of the susceptibility is considered one of the characteristics
of second-order phase transitions. However, it has been pointed out that 
in spin-crossover type systems which belong to the mean-field universality class, 
the spin configuration is uniform even at the critical point. 
Spin-crossover (SC) materials are molecular crystals, 
in which the molecules can exist in two different states: 
the high-spin (HS) state and the low-spin (LS) state. 
The HS state is preferable at
high temperatures because of its high degeneracy, while the LS state is 
preferable at low temperatures because 
it has a low enthalpy.\cite{SC0} 
This type of competition exists not only in SC materials, but also in
charge-transfer materials, Prussian-blue type materials, Jahn-Teller systems, and martensitic materials.
Such systems are generally characterized by the following parameters: the enthalpy difference between the HS and LS states,
the difference between their degeneracies (or entropies), and the strength of 
the intermolecular interactions.
A general classification of types of ordering processes in such systems 
has recently been proposed.\cite{Miya0} 

An important characteristic of this phase transition is an effective long-range
interaction caused by an elastic interaction due to the lattice distortion caused
 by the different sizes of the HS (large) and LS (small) molecules, 
and the spin configuration at the critical point is uniform with 
no large-scale clustering.\cite{Miya1} 
It has also been found that the long-range interaction affects dynamical properties.\cite{Miya2,Mori1} 
In particular, the critical spinodal phenomena predicted by the mean-field theory are truly
realized. This contrasts sharply with the case of short-range models, 
in which the spinodal phenomena occur as
a crossover because nucleation-type fluctuations smear out the 
criticality.\cite{Rikvold}

This uniform spin configuration is one of the crucial characteristics of the pure elastic model without short-range interactions.
However, in real materials, we expect that both short-range and long-range interactions should exist. 
For example, if we consider a usual Lennard-Jones potential between molecules which depends on the spin states,
the model has both elastic and short-range interactions.\cite{Nicolazzi}
In such systems we expect to see ordering clusters due to the short-range interaction, 
though the critical phenomena would still be governed by the long-range interaction.
Thus, it is an interesting problem to study the crossover between short-range and long-range models.\cite{Suzuki} In particular, we expect that the correlation length of the spin-correlation 
function is finite in the thermodynamic limit, even at the critical point, as 
long as any long-range interaction exists.
In the present paper, we study how the critical correlation length increases and ultimately diverges when the 
long-range interaction vanishes.  

To grasp the general features of the competition between the long-range and short-range interactions, 
in this paper
we study a model in which the long-range interactions are those of the Husimi-Temperley model, 
which is the simplest model in which one can study this effect.
We investigate how the spin-correlation function develops due to the short-range interactions.
If the long-range interaction is weak, the system shows ordered clusters near the critical point
of the short-range model, $T_{\rm c}^{\rm IS}$.
If we take the cluster size as the unit length, the model can be regarded as
a pure long-range model, and it shows the critical properties of the mean-field universality
class at the critical temperature of the model, $T_{\rm c}$ . 
This picture enables a scaling analysis of the crossover.
The difference of the critical temperatures,
$T_{\rm c}-T_{\rm c}^{\rm IS}$, is a function of the strength of the long-range interaction.
We introduce a formula for the critical temperature as a function of the 
strength of the long-range interaction, and we perform Monte Carlo 
simulations to confirm this relation. 
A model very similar to ours was previously studied by Hastings as a 
gwell-stirredh approximation for an Ising model on a small-world network.\cite{Hast03}

A characteristic of the present model is that the correlation length is finite, even at the critical point. 
We study how the cluster size diverges as the strength of the long-range interaction 
decreases, and we propose a scaling form for the divergence, which is also confirmed 
by Monte Carlo simulations.

\section{Model : Ferromagnetic Ising model with nearest-neighbor and weak infinite-range interactions}
\label{sec:husimi}

\subsection{Hamiltonian}
First, we consider the effects of a weak, infinitely long-range interaction (Husimi-Temperley model)
on the Ising model with ferromagnetic nearest-neighbor interactions on a square lattice, 
\beq
{\cal H}_{\rm IS}=-J\sum_{\langle i,j \rangle  }\sigma_i\sigma_j,
\label{hamIS}
\eeq
where $ \langle  i,j \rangle $ denotes nearest-neighbor pairs, and $\sigma_i=\pm 1$ denotes the Ising spin
on lattice site $i$.
The critical temperature of this model\cite{Onsager} is
\beq
T_{\rm c}^{\rm IS}= {2J\over \ln(1+\sqrt{2})}\simeq 2.269\cdots J.
\eeq 
We adopt the following Hamiltonian for the long-range interaction: 
\beq
{\cal H}_{\rm HT}=-{4J_0\over 2N}\sum_{i=1}^N\sum_{j=1}^N\sigma_i\sigma_j.
\eeq
The critical temperature of this model\cite{Husimi,Temperley} is
\beq
T_{\rm c}^{\rm HT}=4J_0.
\label{eq-tcHT}
\eeq 
With $J_0=J$, this critical temperature is equal to that of the mean-field approximation for the Ising model
on the present lattice.

For the crossover, we study the following hybrid model,
\beq
{\cal H} = (1-\alpha){\cal H}_{\rm IS}+\alpha{\cal H}_{\rm HT},
\quad 0\leq \alpha \leq 1 \;.
\label{HT}
\eeq
Here, $\alpha$ controls the relative strength of the long-range interaction.

\subsection{Dependence of the critical temperature on $\alpha$}

The critical temperature of the model defined by (\ref{HT}) changes from $T_{\rm c}^{\rm IS}$ to
$T_{\rm c}^{\rm HT}$ as $\alpha$ changes from 0 to 1. 
First, we consider the situation in a naive picture.
At a temperature $T$, the short-range order is developed by ${\cal H}_{\rm IS}$, 
and we assume that $N_{\rm cluster}$ spins are tightly correlated and behave
as one effective spin.
In this case, we introduce an effective spin $\{\tau_i\}, i=1,\cdots N'=N/N_{\rm cluster}$

\beq
S_i=\sum_{j\in {\rm cluster} \; i}^{N_{\rm cluster}}\sigma_j=N_{\rm cluster}\tau_i, \quad \tau_i=\pm 1.
\label{cluster}
\eeq
Using this effective spin, ${\cal H}_{\rm HT}$ is expressed as
\beq
{\cal H}_{\rm HT}=-{4J_0N_{\rm cluster}^2\over 2N_{\rm cluster} N'}
\sum_{i=1}^{N'}\sum_{j=1}^{N'}\tau_i\tau_j
=-{4J_0N_{\rm cluster}\over 2N'}\sum_{i=1}^{N'}\sum_{j=1}^{N'}\tau_i\tau_j.
\eeq
The short-range part has contributions from interactions at 
the interfaces between clusters,
and ${\cal H}_{\rm IS}$ is given by 
\beq
{\cal H_{\rm IS}}\simeq -{J\sqrt[]{N_{\rm cluster}}}\sum_{ \langle i,j \rangle}^{N'}\tau_i\tau_j.
\eeq
As the clusters grow, the long-range interactions become effectively stronger than 
the short-range interactions, and 
the critical temperature is given by
\beq
T_{\rm c}=4\alpha J_0N_{\rm cluster}.
\eeq 
If we estimate $N_{\rm cluster}$ using the Ising correlation length $\xi^{\rm IS}$, 
which has its origin in the short-range interaction, it can be written as 
\beq
N_{\rm cluster}\simeq (\xi^{\rm IS})^{\gamma \over \nu}=(\xi^{\rm IS})^{2-\eta},
\label{xi2}
\eeq
where $\eta$ is the Ising anomalous dimension and the exponent relations are
\beq
\alpha + 2\beta + \gamma =2,
\eeq
\beq
\gamma = (2-\eta)\nu,
\eeq
and $\alpha=0,\beta=1/8, \gamma=7/4, \nu=1,$ and $ \eta=1/4$ in the two-dimensional  
Ising model.\cite{exponent}
Then, using the relation $\xi^{\rm IS}\propto (T-T_{\rm c}^{\rm IS})^{-\nu}$,

\beq
T_{\rm c}-T_{\rm c}^{\rm IS}\propto \left({4\alpha J_0\over T_{\rm c}}\right)^{1\over \gamma}=\left({4\alpha J_0\over T_{\rm c}}\right)^{4\over 7}.
\label{critical_temp}
\eeq
In case $\alpha$ is very small, $T_{\rm c}\simeq T_{\rm c}^{\rm IS}$, so
\beq
T_{\rm c}-T_{\rm c}^{\rm IS}\propto \left({4\alpha J_0\over T_{\rm c}^{\rm IS}}\right)^{4\over 7}\propto \alpha^{4\over 7}.
\label{block_spin}
\eeq
This result agrees with Eq.~(9) of Hastings' paper.\cite{Hast03}
We can confirm the above picture by an exact argument involving the free energy. 
Let us consider the free energy of the total system with a fixed magnetization, $m=\sum_i\sigma_i/N$.
The partition function is given explicitly by
\beq
Z(\beta,m)={\rm Tr}e^{-\beta((1-\alpha){\cal H}_{\rm IS}+\alpha{\cal H}_{\rm HT})}
={\rm Tr}e^{-\beta(1-\alpha){\cal H}_{\rm IS}+\beta 4 \alpha J_0m^2N/2}
=Z^{\rm IS}(\beta(1-\alpha),m)e^{\beta 4 \alpha J_0m^2N/2},
\eeq
where $Z^{\rm IS}(\beta,m)$ is the partition function of the Ising model
at the inverse temperature $\beta$ for a fixed value of $m$.
Therefore, the free energy is given by
\beq
F(\beta,m)/N=-{1\over \beta N}\ln(Z(\beta,m))=(1-\alpha)f^{\rm IS}(\beta(1-\alpha),m)-4\alpha J_0m^2/2.
\eeq 
Here $f^{\rm IS}$ is the free energy per spin of the Ising model, 
and we assume that it can be expanded around the
critical point in the following form
\beq
f^{\rm IS}(\beta(1-\alpha), m)\simeq {1\over 2\chi^{\rm IS}(\beta(1-\alpha))
}m^2+\cdots.
\eeq
Thus the critical point of the present model is given by ${\partial^2 F(\beta_{\rm c},m) \over \partial m^2}|_{m=0}=0$, or 
\beq
{(1-\alpha)\over \chi^{\rm IS}(\beta_{\rm c}(1-\alpha))}-{4\alpha J_0}=0,
\label{eq-tcalpha}
\eeq
where the susceptibility of the hybrid model (\ref{HT}) diverges. If we adopt the relation
\beq
\chi^{\rm IS}(T)\propto (T-T_{\rm c}^{\rm IS})^{-\gamma},
\label{IS_sus}
\eeq
the critical point is given by
\beq
T_{\rm c}-(1-\alpha)T_{\rm c}^{\rm IS}\propto(4\alpha J_0)^{1\over \gamma}(1-\alpha)^{1-{1\over\gamma}}\simeq \alpha^{1\over \gamma}
\label{critic_temp}
\eeq
which agrees with (\ref{block_spin}) for small $\alpha$. 
We note that equation~(\ref{IS_sus}) holds only when $T$ is very close to $T_{\rm c}^{\rm IS}$, 
so equation~(\ref{critic_temp}) holds only when $T_{\rm c}$ is very close to $(1-\alpha)T_{\rm c}^{\rm IS}$. 
Namely, equation~(\ref{critic_temp}) is only valid for $\alpha \ll 1$.

For $\alpha =1$, 
\beq
\lim_{\alpha \rightarrow 1}F(\beta,m)/N=\lim_{\alpha \rightarrow 1}(1-\alpha)f^{\rm IS}(\beta(1-\alpha),m)-4\alpha J_0m^2/2={T \over 2}m^2-4J_0m^2/2+\cdots \quad .
\eeq
In this case (\ref{eq-tcalpha}) yields $T_c = 4J_0$, the critical temperature 
of the Husimi-Temperly model (\ref{eq-tcHT}).

To obtain the numerically correct amplitude for $T_{\rm c}(\alpha)$, we need the Ising susceptibility 
near the critical point for $T>T_{\rm c}^{\rm IS}$,\cite{ising_sus}
\beq
\chi^{\rm IS}(\beta)=\beta C_{\rm 0}\left({1\over \tilde{t}}\right)^{\gamma},\hspace{0.5cm}\gamma=7/4
\eeq
with $C_{\rm 0}=0.962582\cdots$, and
\beq
\tilde{t}={T-T_{\rm c}^{\rm IS}\over T}={T_{\rm c}^{\rm IS}\over T}t.
\eeq
Equation (\ref{eq-tcalpha}) can be written as
\beq
{T_{\rm c}\over C_{\rm 0}}\left({(1-\alpha)T_{\rm c}^{\rm IS}\over T_{\rm c}}\right)^
{7\over 4}\left({T_{\rm c}-(1-\alpha)T_{\rm c}^{\rm IS}\over(1-\alpha)T_{\rm c}^{\rm IS}}\right)
^{7\over 4}=4\alpha J_0.
\eeq
We write $t_{\rm c}(\alpha)={T_{\rm c}-T_{\rm c}^{\rm IS}\over T_{\rm c}^{\rm IS}}$,
so
\beq
(1+t_{\rm c})^{4\over 7}(t_{\rm c}+\alpha)=(1+t_{\rm c})
\left({4\alpha J_0 C_0\over T_{\rm c}^{\rm IS}}\right)^{4\over 7}.
\eeq
Expanding to lowest order in $t_{\rm c}$ and $\alpha$ while setting $J_0=J$, 
we get 
\beq
t_{\rm c}\simeq A\alpha^{4\over 7} \hspace{0.25cm}{\rm with}\hspace{0.25cm}A=\left({4J_0C_0\over T_{\rm c}^{\rm IS}}\right)^{4\over 7}\simeq 1.352745,
\eeq
or equivalently,
\beq
{T_{\rm c}(\alpha)-T_{\rm c}^{\rm IS}\over T_{\rm c}^{\rm MF}-T_{\rm c}^{\rm IS}}\simeq 1.773517\alpha^{4\over 7}.
\label{exact_tc}
\eeq
This result agrees with (\ref{block_spin}). 

\subsection{Monte Carlo study of the $\alpha$ dependence of $T_{\rm c}$}
\label{sec:husimiMC}

In order to confirm the scaling relation of the previous subsection,
we estimated the critical temperatures for various values of $\alpha$
by Monte Carlo simulations. 
Here we fixed both $J$ and $J_0$ to 1.0.
Therefore, $T_c^{\rm IS}=2.269\cdots$ and $T_c^{\rm HT}=4$ in these units.
We used a standard Metropolis method, adopting periodic boundary conditions. 
In most cases, we performed 500,000 MCS (Monte Carlo steps) for the data with 100,000 MCS for the equilibration. 
From now on, $L$ denotes the linear system size in units of the lattice constant, so the total number of spins is $L^2$.

\begin{figure}
\includegraphics[scale=0.4]{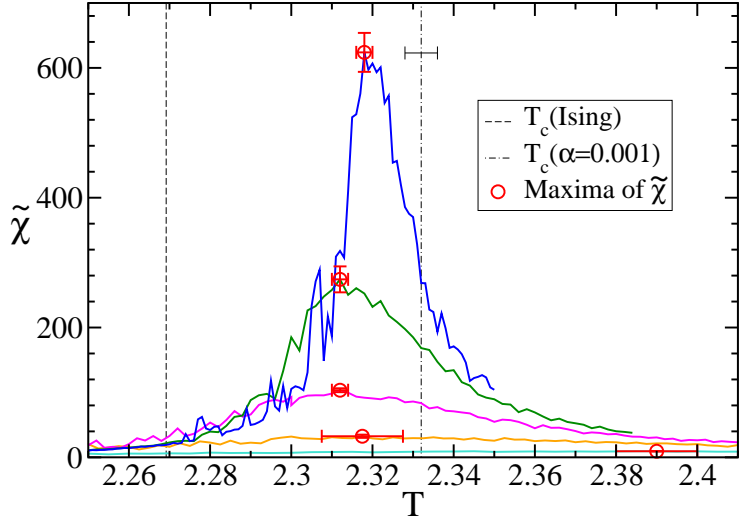} 
\caption{(Color online) Temperature dependences of $\tilde{\chi}$ for $\alpha=0.001$ and $L=20, 40, 80, 160,$ and $320$ from below to above. 
The peak positions are marked by circles. 
For large systems, they increase with increasing $L$. The left vertical dashed line represents the critical temperature for the pure Ising model, 
and the right-hand line represents the critical temperature for $\alpha=0.001$ obtained by the 
Binder cumulant method (see below).}
\label{chipeak}
\end{figure}

We estimate a candidate for the critical temperature $T_{\rm c}(\alpha)$ for each value of $\alpha$.
In order to obtain this value, we study the size dependence of the peak position of 
the so-called absolute susceptibility\cite{Landau}
\beq
\tilde{\chi}\equiv {1\over N}(\langle M^2 \rangle - \langle |M| \rangle^2)
\eeq
as a `critical point' $T_{\rm c}(\alpha,L)$ for the size $L$.
We expect that the peak position saturates at the critical temperature 
in the thermodynamic limit:
\beq
T_{\rm c}(\alpha,L)\rightarrow T_{\rm c}(\alpha,\infty).
\eeq
In Fig. \ref{chipeak}, we depict a typical size dependence of the peak for $\alpha=0.001$.
By a general argument we expect the following size dependence:
\beq
T_{\rm c}(\alpha,L)-T_{\rm c}(\alpha,\infty)\propto L^{-1/\nu}.
\label{TcL0}
\eeq
Here, $\nu$ is the critical exponent of the correlation length.
However, in the present case, the critical phenomena belong to the mean-field 
universality class, and the definition of $\nu$ is subtle.
Namely, if we consider the spatial correlation of the Gaussian model,
$\nu=1/2$, while
in the scaling relation in the mean-field universality class, 
we have the {\it effective} $\nu=2/d$.\cite{Miya1,privman,Luijten1} 
In the infinite-range (HT) model, distances are not well defined, and 
only the total number of spins,  $N$, has
a meaning. Thus, we should rewrite the relation (\ref{TcL0}) as
\beq
T_{\rm c}(\alpha,L)-T_{\rm c}(\alpha,\infty)\propto N^{-1/d\nu},
\eeq
where we take the latter case ($\nu=2/d$) as we did in a previous 
paper.\cite{Miya1}
In the present case $d=2$ and thus $N=L^2$, which gives
\beq
\nu_{\rm HT}=1,
\eeq
which accidentally agrees with that of the short-range Ising model,
\beq
\nu_{\rm IS}=1.
\eeq

\begin{figure}
$$
\includegraphics[scale=0.4]{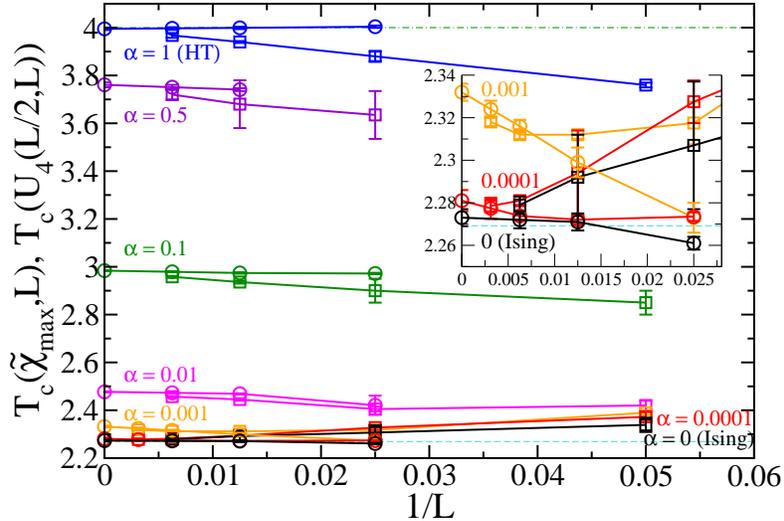}
$$
\caption{(Color online) Estimates for $T_{\rm c}(\alpha,L)$ for different values of $\alpha$ vs $L^{-1}$. 
$\alpha=0$ (pure Ising), $0.0001, 0.001, 0.01, 0.1, 0.5,$ and $1$ (Husimi-Temperley) from below to above. 
$L=20, 40, 80, 160,$ and $320$. 
The squares denote the peak positions of $\tilde{\chi}$, and the circles 
denote
the crossing positions of the Binder cumulant for $L$ and $L/2$.
The upper and lower dashed lines mark the exact critical temperatures for 
the HT and Ising models, respectively. 
The inset shows detail for $\alpha=0.001, 0.0001,$ and $0$ (pure Ising).
}
\label{TCsize}
\end{figure}

In Fig.\ \ref{TCsize}, we plot the peak position of $\tilde{\chi}$ by open 
squares as a function of $L^{-1}$ for several values of $\alpha$. 
The critical temperature $T_{\rm c}(\alpha,\infty)$ could in principle 
be estimated 
by linear extrapolation in $L^{-1}$.  
However, we find a non-monotonic dependence of the peak position as a function of $L^{-1}$
for small values of $\alpha$. (See also Fig.~\ref{chipeak}.) 
Only when the size becomes large enough to show 
the critical behavior of the HT model, can we apply the 
scaling relation (\ref{TcL0}).
For small sizes, the system behaves like a short-range model, and the peak 
position moves differently. Indeed, we find that in the scaling region, 
the peak position approaches $T_c(\alpha,\infty)$ from below. 
However, for $\alpha \leq 0.01$ we find that 
it decreases with $L$ for small values of $L$.
For $\alpha=0.001$, we find that the peak position 
finally increases again when $L$ goes from $160$ to $320$, while
for $\alpha=0.0001$, it continues to decrease for all values of $L$ considered.
Thus, we cannot estimate the infinite-system value by a simple 
extrapolation of the peak position in $L^{-1}$ for $\alpha=0.0001$.

\begin{figure}[t]
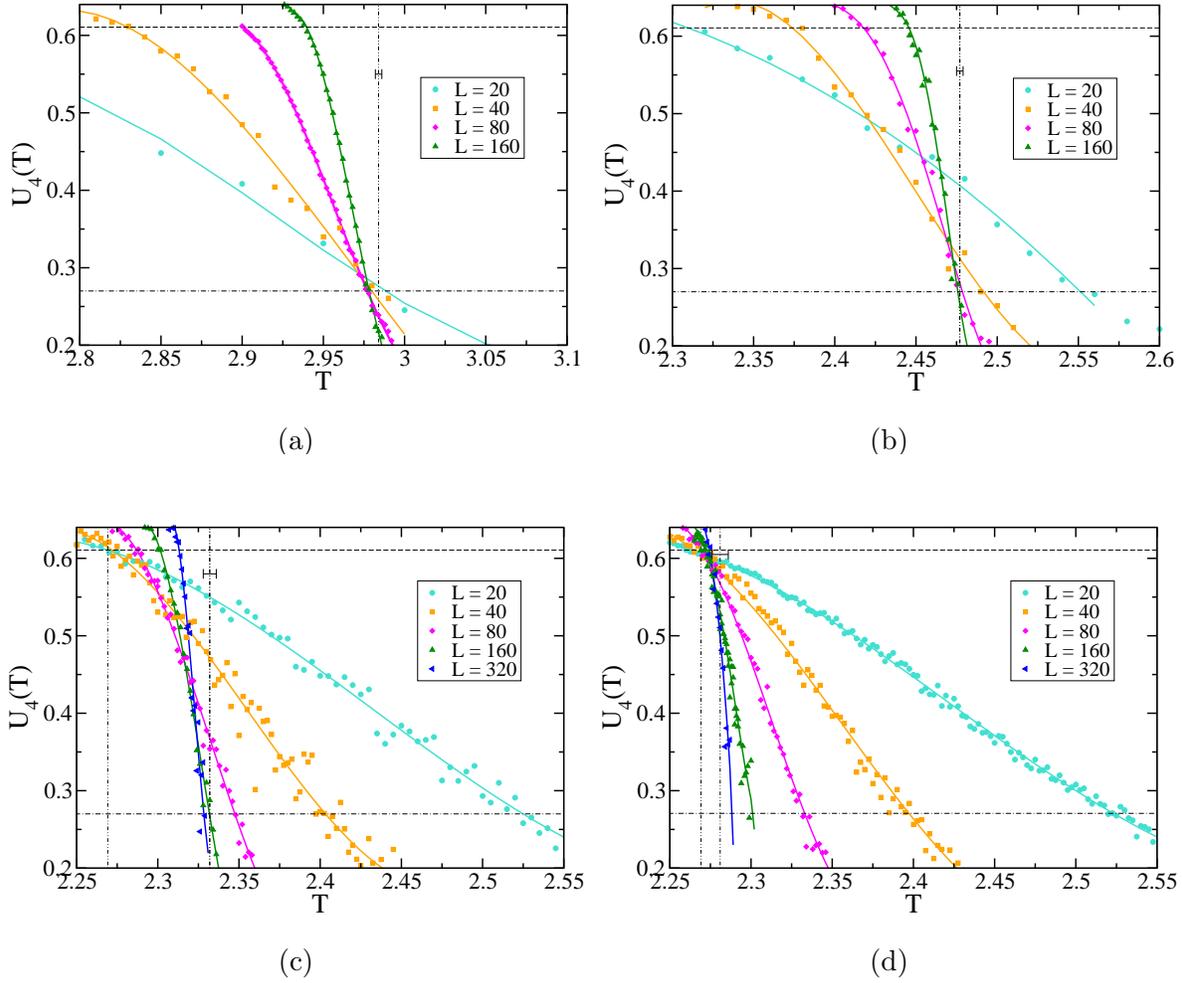

$$\begin{array}{cc}
\includegraphics[scale=0.3]{03Mixed01Mlongrun_Bx_detail_2_v2.eps} &
\includegraphics[scale=0.3]{04Mixed001Mlongrun_Bx_detail_2_v2.eps} \\

({\rm a}) & ({\rm b})\\
\\

\includegraphics[scale=0.3]{05Mixed0001Mlongrun_Bx_detail_v2.eps} &
\includegraphics[scale=0.3]{06Mixed00001Mlongrun_Bx_detail_v2.eps} \\
({\rm c}) & ({\rm d})
\end{array}
$$
\caption{(Color online) Temperature dependence of the Binder cumulant for
(a) $\alpha=0.1$, (b) $\alpha=0.01$, (c) $\alpha=0.001$, and (d) 
$\alpha=0.0001$.
Points are Monte Carlo data, and the solid lines are polynomial fits. 
The upper and lower horizontal lines are the fixed-point values for the Ising 
model and the Husimi-Temperley model respectively, and the left vertical lines in (c) and (d) represent 
the critical temperature of the pure Ising model. 
The vertical dashed 
lines with horizontal error bars represent the critical temperatures obtained by extrapolation of the 
crossing temperatures as described in Appendix \ref{appendixA}. 
}
\label{Binderalpha} 
\end{figure}

To obtain more accurate estimates for small $\alpha$, 
we also estimated $T_c(\alpha,L)$ from the crossing point of the 
fourth-order Binder cumulant,\cite{bindercumulant} 
\beq
U_4(\alpha,L) = 1 - {\langle m^4 \rangle \over 3 \langle m^2 \rangle^2},
\label{eq-Binder}
\eeq
for $L$ and $L/2$. 
When $L$ and $\alpha$ are small, the crossing value of 
$U_4(\alpha,L)$ is near the Ising fixed-point value, 
$U_4^{*\rm IS}\simeq 0.61...$,\cite{IsingCum} 
while for larger $L$ and/or $\alpha$, the crossing moves down toward the exact 
value for  the HT model, 
$U_4^{*\rm HT}= 1 - \Gamma^4(1/4)/24 \pi^2 = 0.27...$,\cite{Brezin,Luijten} 
where $\Gamma(x)$ is the Gamma function. 
The values of $T$ at the crossing points are shown as circles vs $L^{-1}$ 
for different values of $\alpha$ in Fig.~\ref{TCsize}. 
The temperature dependences of $U_4(\alpha,L)$ for different $\alpha$ and $L$
are shown in Fig.~\ref{Binderalpha}.
For $\alpha=0.1$, we find the crossing points located near $U_4^{*\rm HT}$, 
indicating that
the critical properties belong to the mean-field universality class.
For $\alpha=0.01$ and 0.001, we find that the crossing points move from 
near $U_4^{*\rm IS}$ toward $U_4^{*\rm HT}$ as $L$ increases.
These results indicate that
the critical point of the hybrid model belongs to the mean-field universality 
class for all $\alpha > 0$.
In the case of $\alpha=0.0001$, the crossing point of the Binder cumulants 
for $(L,L/2)=(320,160)$ is still near $U_4^{*\rm IS}$.
Because we assume that the critical behavior for nonzero $\alpha$ belongs 
to the mean-field 
universality class, we get a series of upper bounds on the critical temperature as 
the temperature at which 
$U_4(\alpha,L)$ crosses $U_4^{\rm HT}$. Lower bounds are given by the 
cumulant-crossing temperatures. Our best estimates for $T_c$ are obtained by 
linearly extrapolating the crossing temperatures to $L^{-1} = 0$ . 
In this way, we estimated the 
$T_{\rm c}(\alpha=0.0001)=2.281\pm 0.005$.
In Appendix \ref{appendixA} we show in detail how we estimated this value.

\begin{figure}[ht]
$$
\includegraphics[scale=0.4]{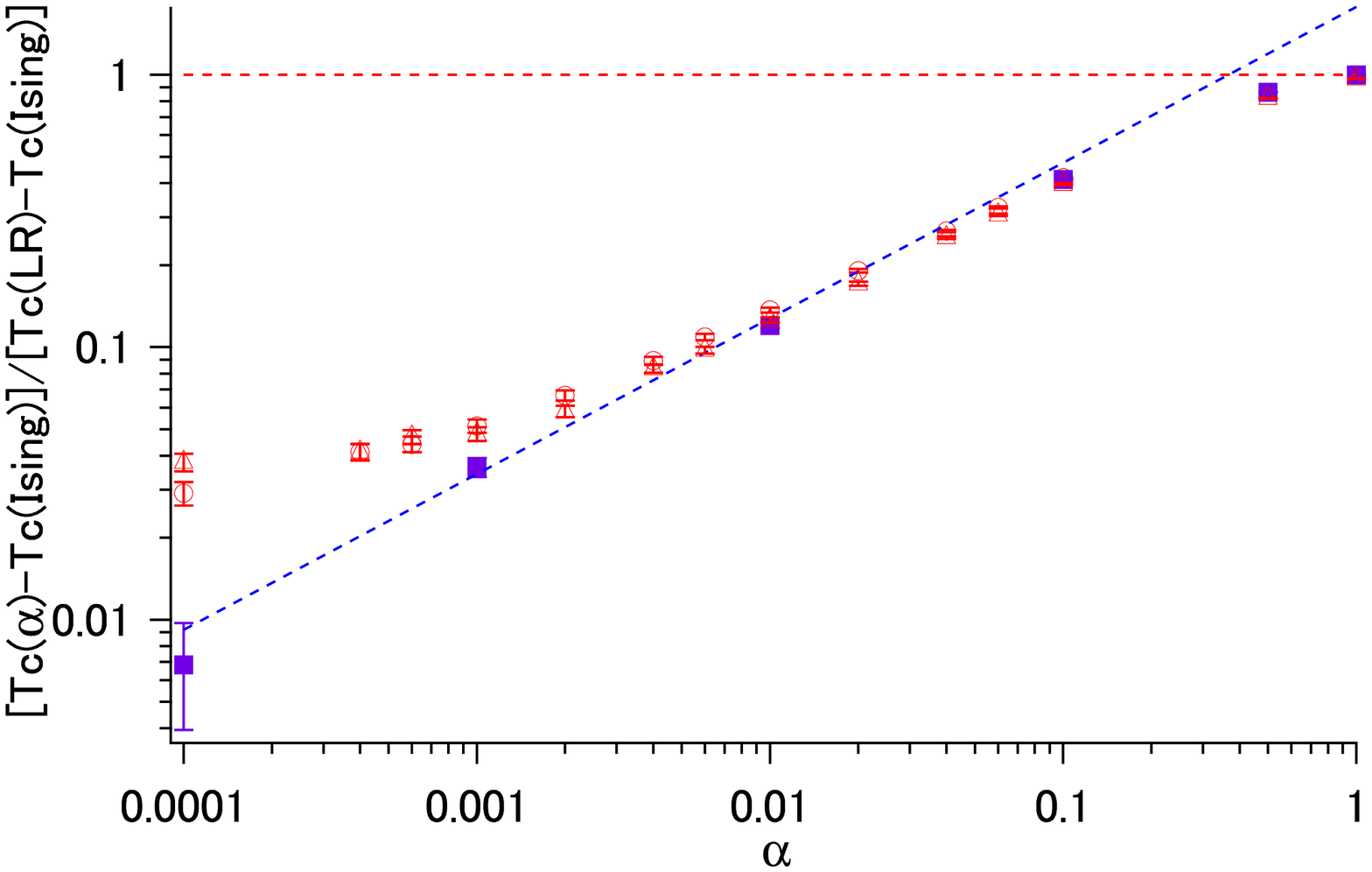}%
$$
\caption{(Color online) The $\alpha$ dependence of the normalized critical-temperature 
difference, 
$[T_{\rm c}(\alpha)-T_{\rm c}^{\rm IS}]/[T_{\rm c}^{\rm HT}-T_{\rm c}^{\rm IS}]$
in a log-log plot.
 The circles and triangles (red online) denote critical temperatures 
obtained from the peak position of $\tilde{\chi}$ for $L=320$ and $80$ respectively,
 and the squares (blue online) denote critical temperatures obtained from Binder cumulants. 
The horizontal dashed line (red online) represents $y=1$, 
and the oblique dashed line (blue online) represents the numerically exact theoretical estimate, 
$1.773517 \alpha^{4/7}$ (\ref{exact_tc}). 
The latter line, which involves {\it no} adjustable parameters, agrees very well with the 
cumulant-generated data for small $\alpha$.
}
\label{TCgraph} 
\end{figure}

The extrapolated values for $T_c(\alpha,\infty)-T_{\rm c}^{\rm IS}$ are shown 
on a log-log scale in Fig.~\ref{TCgraph}. 
For small $\alpha$, the data points fall on a straight 
line of slope ${4/7}$. 
As mentioned above, to obtain more accurate estimates of the critical temperatures from $\tilde{\chi}$, 
we would need to perform MC with much larger systems. 
For small $\alpha$, the results from the Binder cumulants are in good agreement 
with the power law, 
$T_{\rm c}(\alpha)-T_{\rm c}^{\rm IS} \propto \alpha^{4/7}$. 
Thus, we confirm the scaling relation (\ref{critical_temp}):
\beq
T_c(\alpha,\infty)-T_c(0,\infty)\propto \alpha^{1/\gamma_{\rm IS}}.
\eeq

\section{Cluster size at the critical point}

\begin{figure}[htbp]
$$\begin{array}{cc}
\includegraphics[scale=0.4]{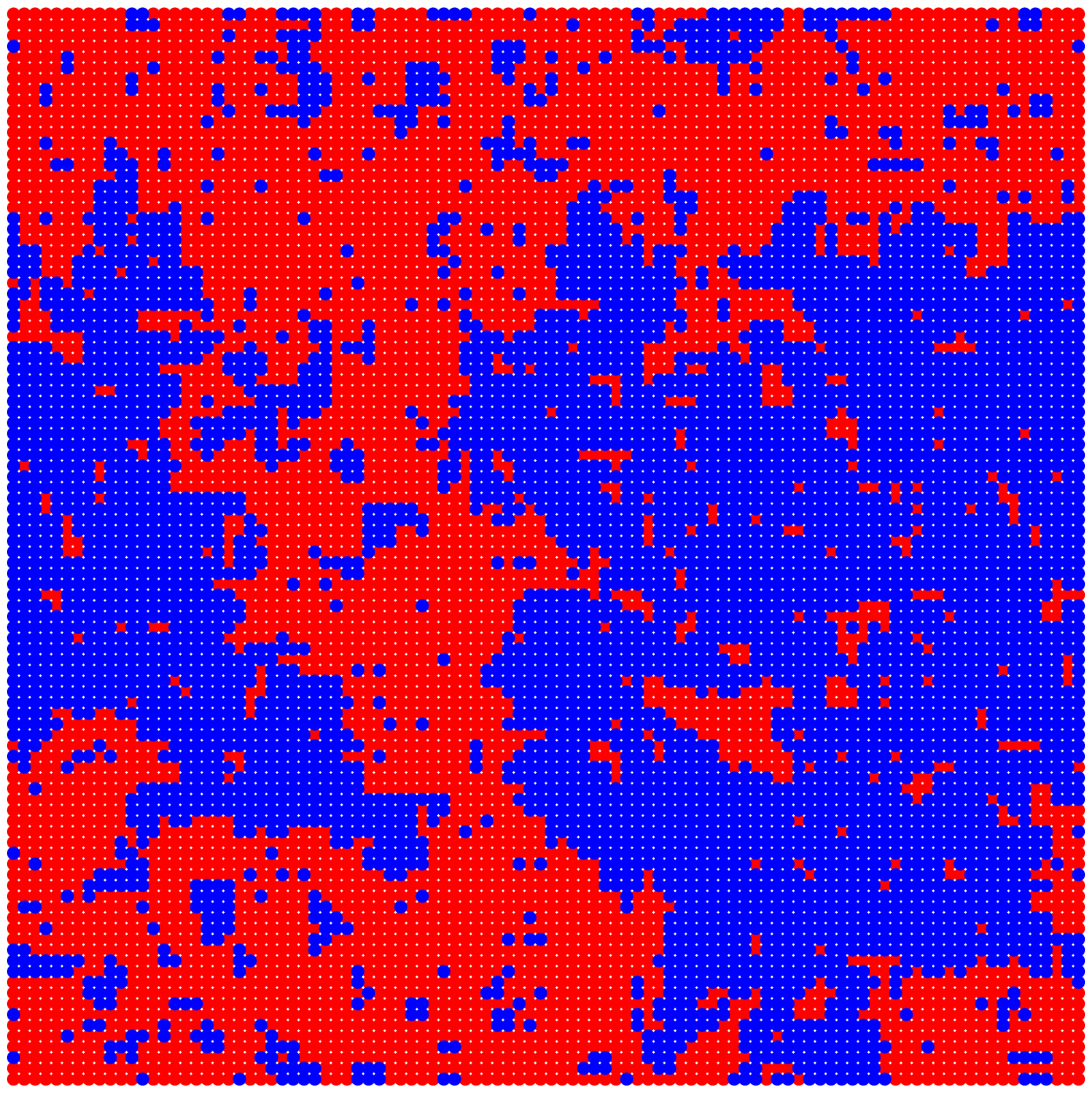}&
\includegraphics[scale=0.4]{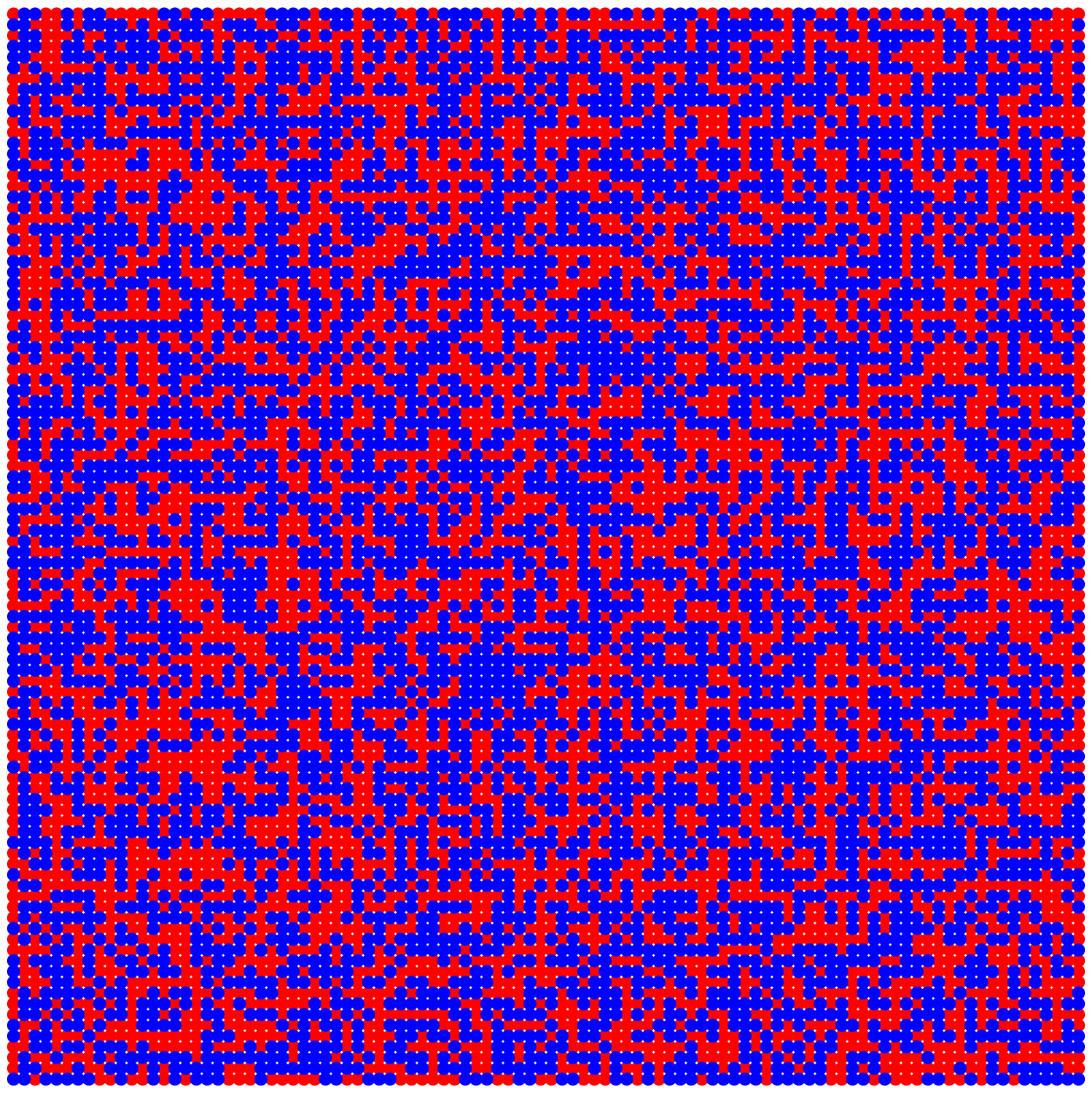}\\
({\rm a}) & ({\rm b}) \end{array}
$$
\caption{(Color online) (a) Spin configurations for Ising model at $T_{\rm c}^{\rm IS} = 2.269J$, $L=100$ 
and (b) Husimi-Temperley model at $T_{\rm c}^{\rm HT}=4J_0$.}
\label{isingsnap}
\end{figure}

In the pure long-range model, all spins interact with each other. 
Thus, the concept of distance has no meaning, and the system does not show any 
clustering.
On the other hand, in the short-range model, the ordering process occurs as a 
development of
short-range order, and the cluster size, i.e., the correlation length, 
represents the extent of the ordering.
In Fig.~\ref{isingsnap}(a) and Fig.~\ref{isingsnap}(b), we depict typical
spin configurations at the critical 
temperature $T_{\rm c}(\alpha)$ of the short-range model and the long-range model, respectively. 
A clear difference between the two cases is evident. 

\begin{figure}[htbp]
$$\begin{array}{ccc}
\includegraphics[scale=0.3]{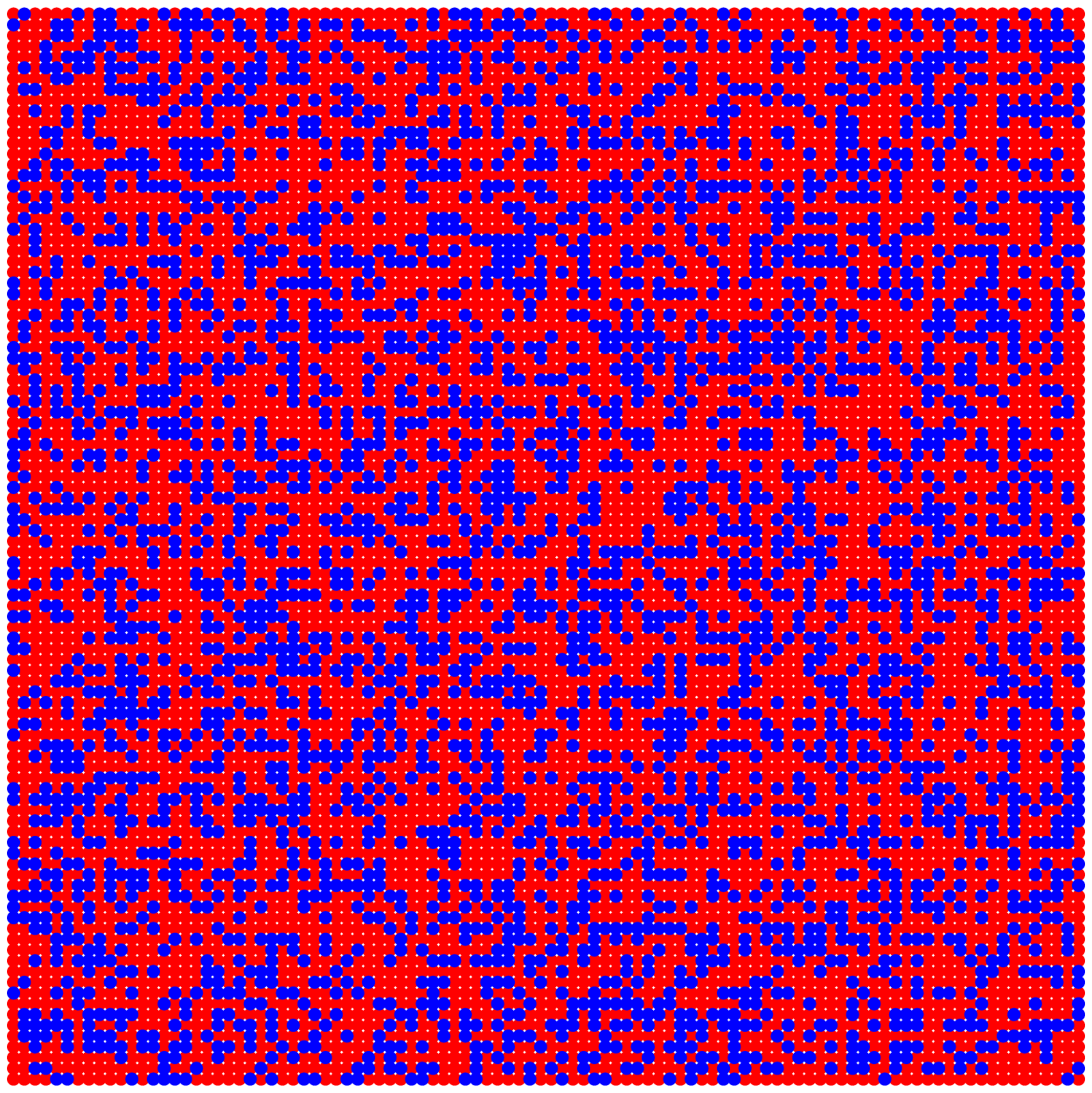}&
\includegraphics[scale=0.3]{09husimi1.eps}&
\includegraphics[scale=0.3]{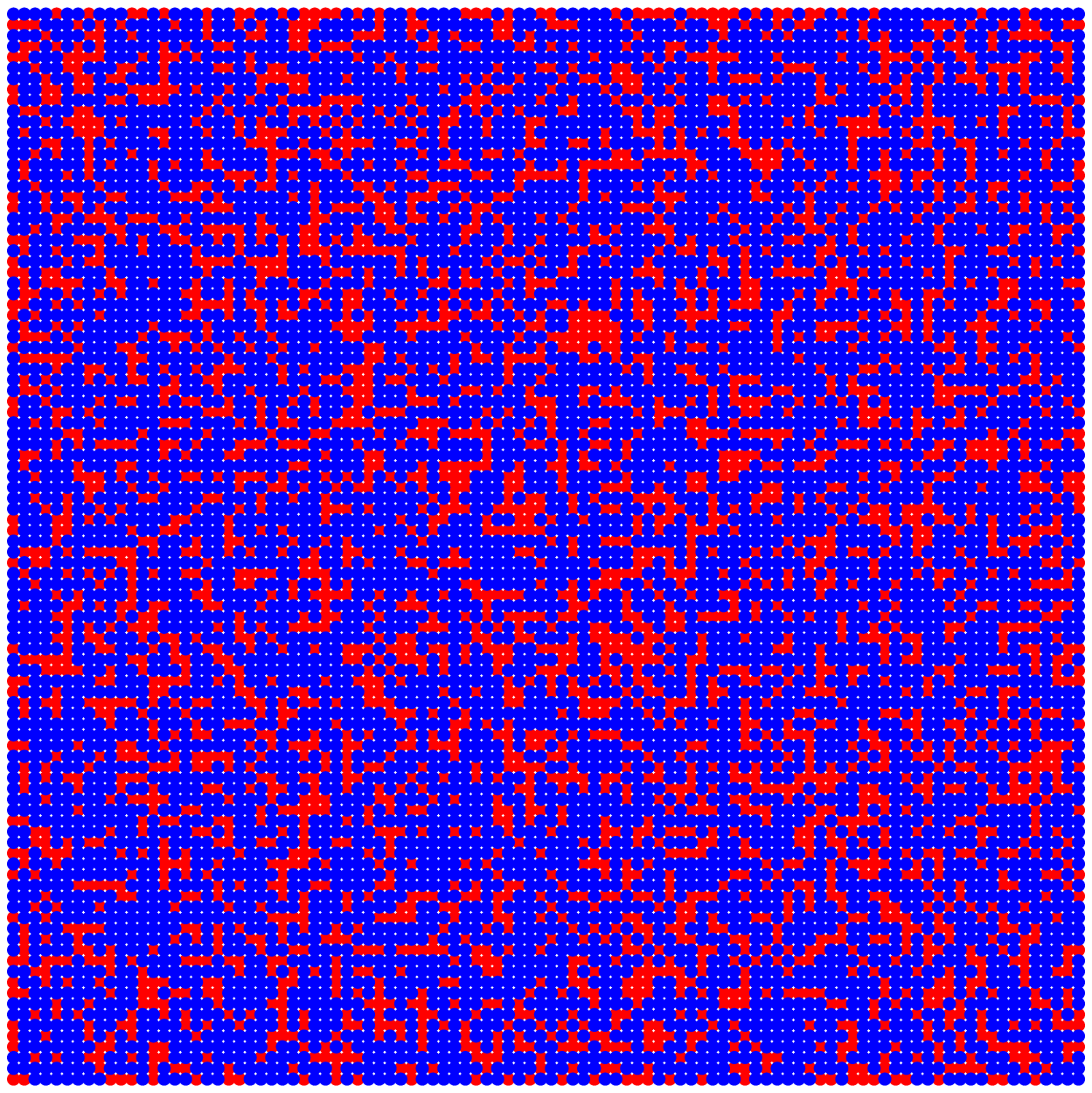}\\
({\rm a}) & ({\rm b})& ({\rm c})  \end{array}
$$
\caption{(Color online) Spin configurations for the Husimi-Temperley model for (a)$\langle m \rangle$ $\simeq$ 0.3 at $T_{\rm c}^{\rm HT}$,
(b)$\langle m \rangle$ $\simeq$ 0.0 at $T_{\rm c}^{\rm HT}$
and (c) $\langle m \rangle$ $\simeq$ -0.3 at $T_{\rm c}^{\rm HT}$.}
\label{husimi_mag}
\end{figure}

Here it should be noted that non-divergence of the correlation length does not mean non-divergence of the susceptibility. In the mean-field model, the 
susceptibility 
diverges as $|T-T_{\rm c}^{\rm HT}|^{-1}$. This means that the fluctuation of 
the magnetization $M$ diverges as 
\beq
{1 \over N}(\langle M^2\rangle - \langle M\rangle^2) \propto |T-T_{\rm c}^{\rm HT}|^{-1}.
\eeq
This fluctuation can be observed as the fluctuation of the uniform density of 
the spin configuration. In Fig.~\ref{husimi_mag}
we depict typical configurations at $T_{\rm c}^{\rm HT}$ with different $M/N=m$. 
We note that the spin configurations are uniform, but the ratio of numbers of 
up and down spins fluctuates. This causes large fluctuations in the 
magnetization $M$, but not in the cluster size.
In the long-range model, large numbers of spins change uniformly.

\begin{figure}[htbp]
$$\begin{array}{cc}
\includegraphics[scale=0.3]{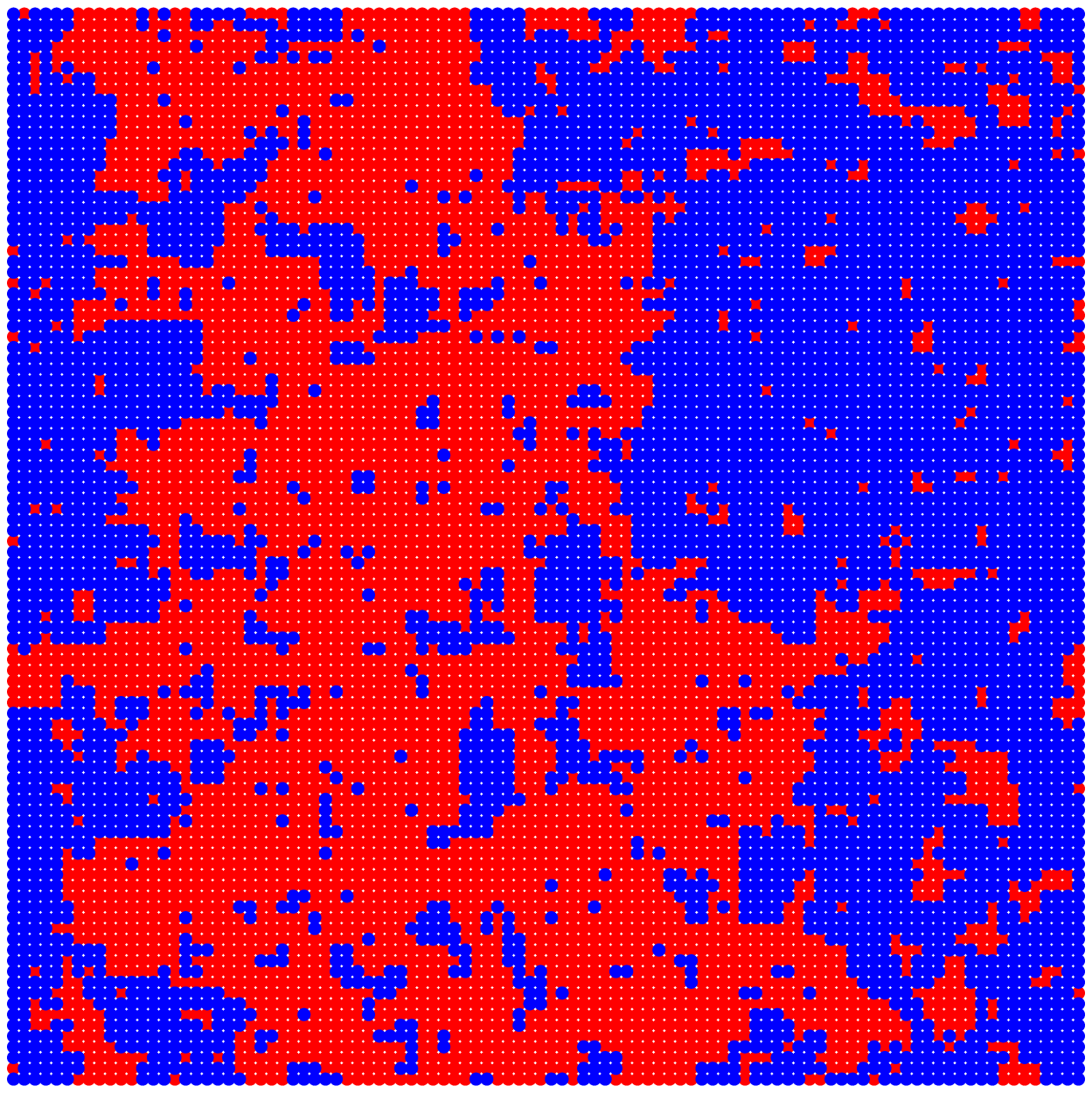} & 
\includegraphics[scale=0.3]{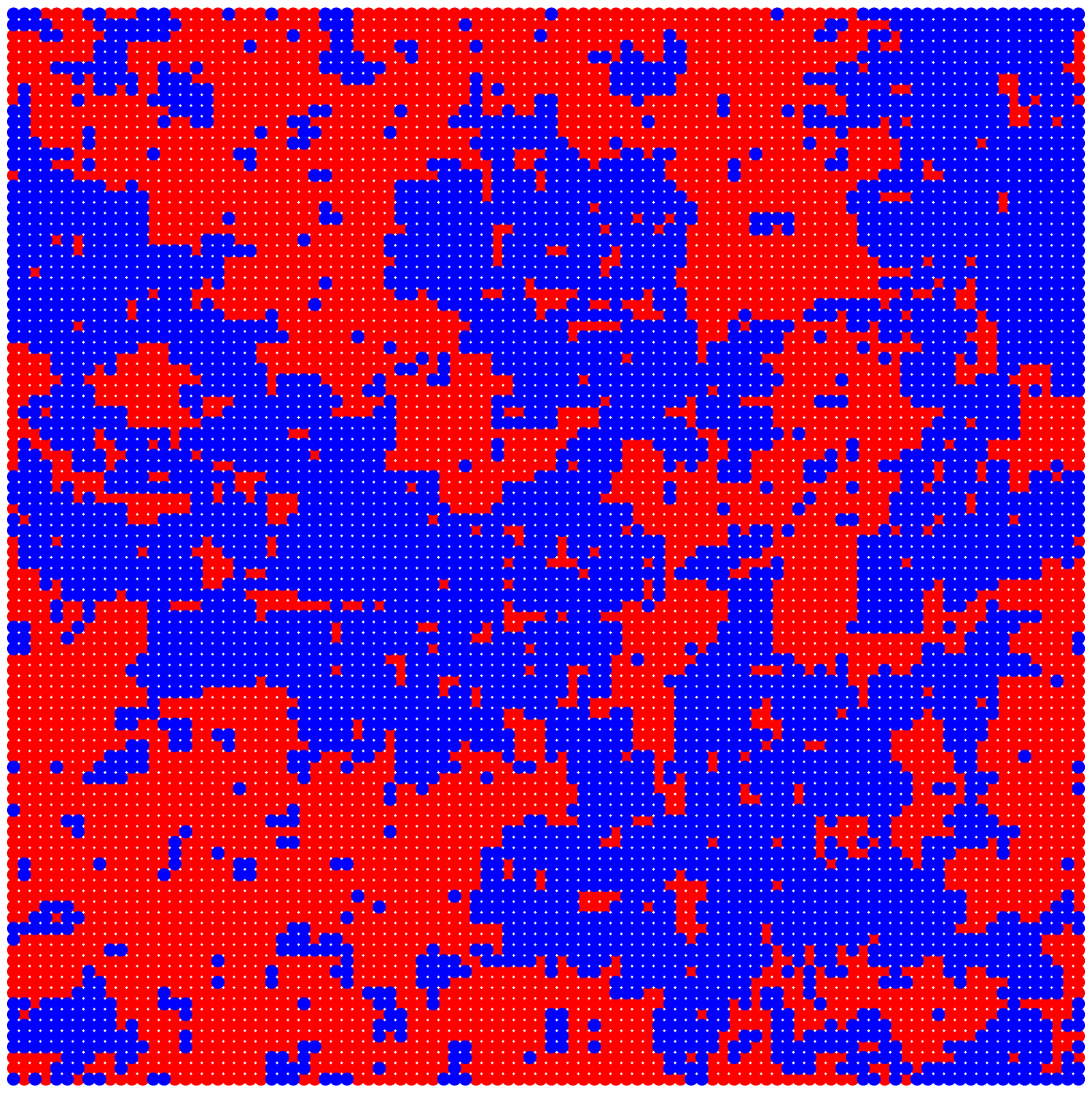}\\
({\rm a}) & ({\rm b})\\
\includegraphics[scale=0.3]{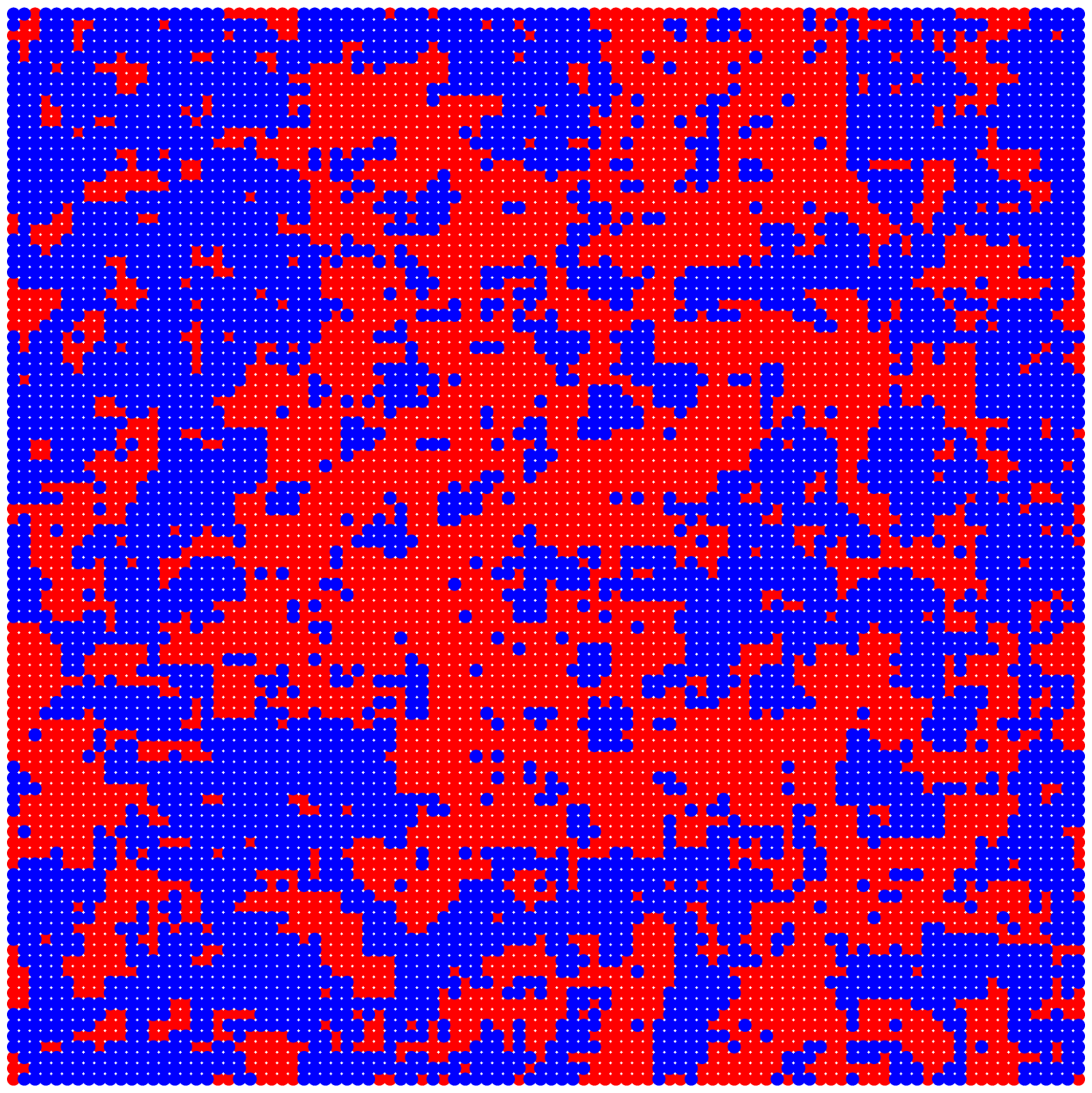} & 
\includegraphics[scale=0.3]{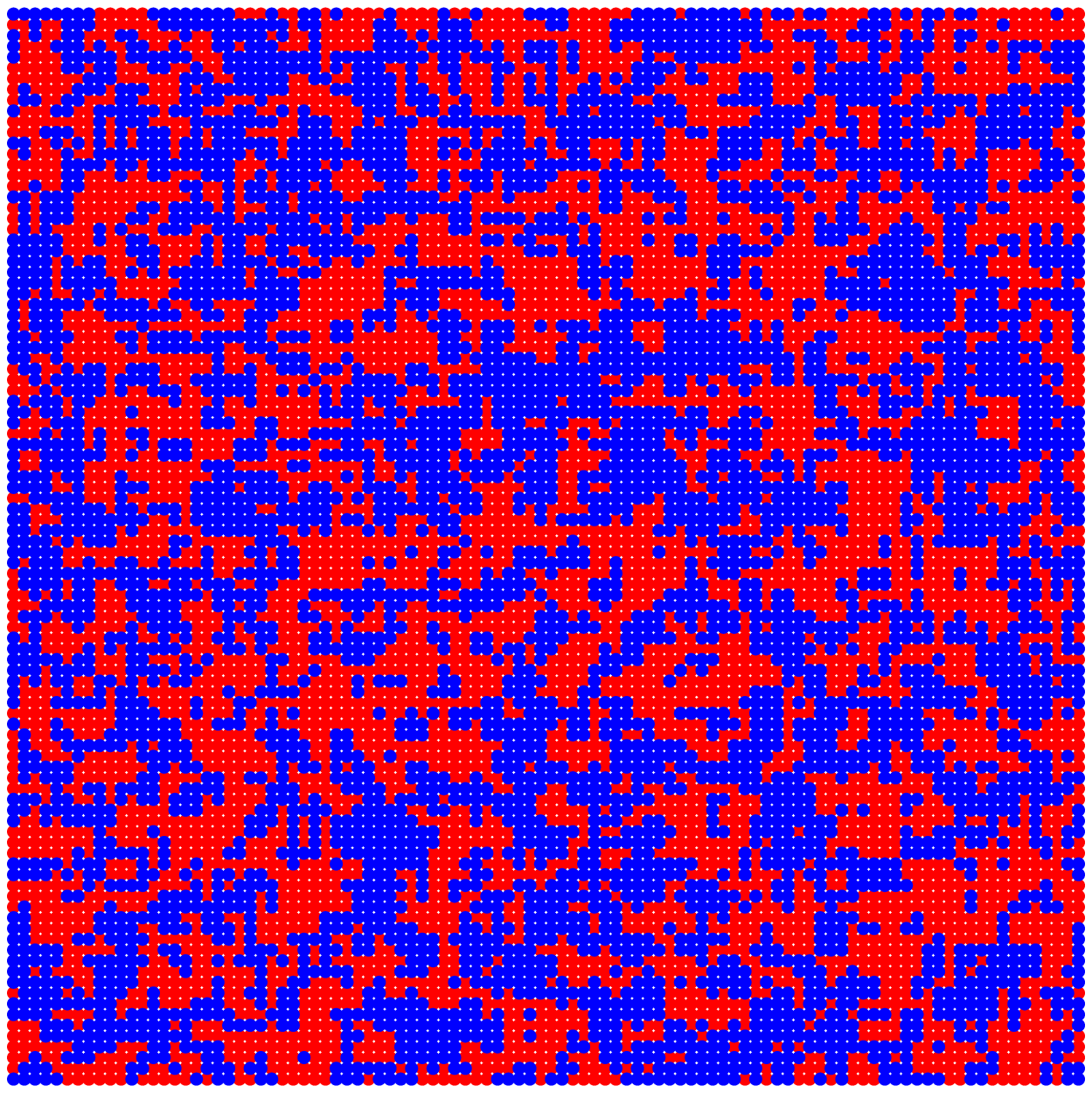}\\
({\rm c}) & ({\rm d})\end{array}
$$
 \caption{(Color online) Typical configurations of the hybrid model at the critical temperature 
$T_{\rm c}(\alpha)$ for (a) $\alpha=0.0001$, (b) $\alpha=0.001$, (c) $\alpha=0.01$, and (d) $\alpha=0.1$.}
\label{Confalpha}
\end{figure}

In the hybrid model (\ref{HT}), the criticality belongs to the mean-field 
universality class.
However, short-range order also develops. Thus, we expect a finite correlation
length at the critical point, which increases as $\alpha$ decreases.
In Fig.~\ref{Confalpha}, we depict typical configurations at the critical 
temperature for various values
of $\alpha$. We clearly see that the size of the clusters increases with 
decreasing $\alpha$.

\section{Finite-size scaling of the cluster size at the critical point}
\label{sec:cluster}

\subsection{Scaling function}
In this section, we study the correlation length at 
the critical point for several values of $\alpha$.
From the relation (\ref{critic_temp}), we expect the following relation between the correlation length $\xi_{\rm c}$ 
at the critical temperature and $\alpha$ :

\beq
\xi_{\rm c}(\alpha) \propto (T_{\rm c}-(1-\alpha)T_{\rm c}^{\rm IS})^{-\nu_{\rm IS}}\simeq
\left(
(4\alpha J_0\chi_0)^{1/\gamma_{\rm IS}}
\right)^{-\nu_{\rm IS}}\propto \alpha^{-\nu_{\rm IS}/\gamma_{\rm IS}}
\eeq
for $\alpha \ll 1$. 
Moreover, for $\alpha \ll 1$ 
we may assume the following finite-size scaling relation with the linear dimension of the system $L$
\beq
\xi_{\rm c}(\alpha, L)=Lf\left(L\alpha^{\nu\over \gamma}
\right)=Lf\left(L\alpha^{4\over 7}
\right),
\label{scalingxialpha}
\eeq
where $f(x)$ is a scaling function which is proportional to $1/x$ for large $x$, and constant for small $x$.

In the case of the short-range Ising model, we can estimate 
the divergence of the correlation length by making use of the susceptibility:
\beq
\chi={1\over Nk_{\rm B}T}\sum_{i,j}\langle\sigma_i\sigma_j\rangle
\sim {1\over k_{\rm B}T}\int_0^L {1\over r^{d-2+\eta}}e^{-r/\xi} dr \sim
\xi^{2-\eta}=\xi^{\gamma \over \nu}.
\eeq
However, in the present long-range interaction model, the value of the magnetization 
fluctuates uniformly but not spatially. Therefore, we cannot estimate 
$\xi$ from $\chi$.

\subsection{Measurement of $\xi$ at $T_{\rm c}$ : Direct measurement of the correlation function}

\begin{figure}
$$
\includegraphics[scale=0.4]{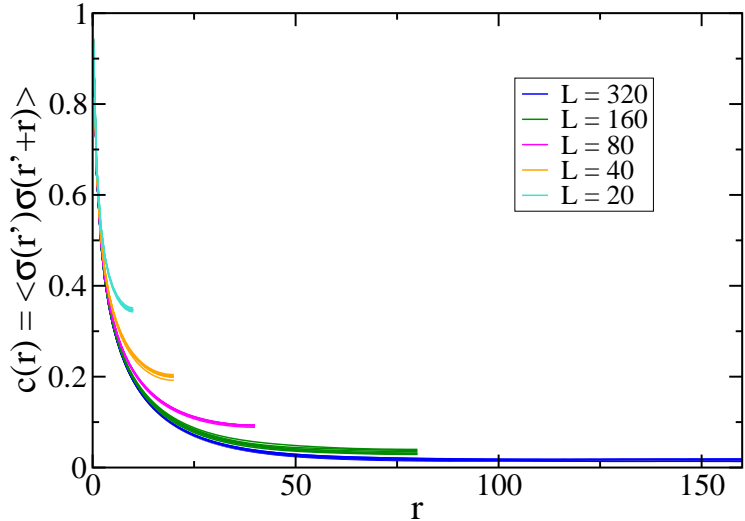}
$$
\caption{(Color online) The disconnected spin correlation function 
$c(r)$ at the critical point $T_{\rm c}(\alpha)=2.332$ for $\alpha=0.001$. 
For each value of $L$, the results of seven independent runs of $10^6$ MCS each are shown.}
\label{Xi}
\end{figure}

\begin{figure}
$$
\includegraphics[scale=0.4]{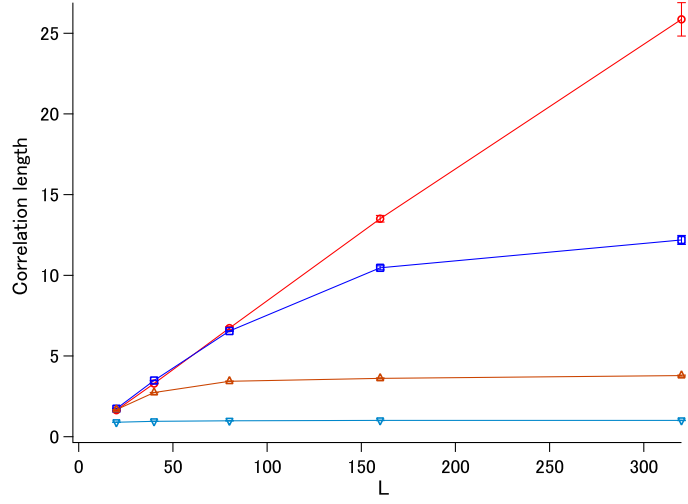}
$$
\caption{(Color online) The size dependence of the correlation length 
at the critical point $T_{\rm c}(\alpha)$. Circles, squares, up-triangles, and down-triangles represent $T=2.281$ for $\alpha=0.0001$, 
$T=2.332$ for $\alpha=0.001$, $T=2.477$ for $\alpha=0.01$ and $T=2.984$ for $\alpha=0.1$, respectively.
}
\label{XialphaL}
\end{figure}

\begin{figure}
$$
\includegraphics[scale=0.4]{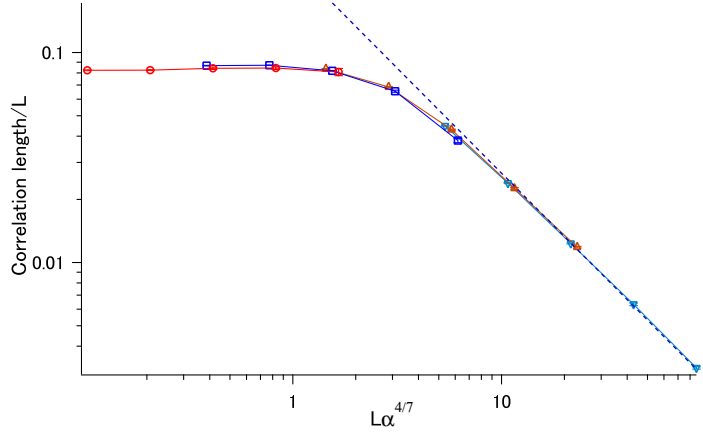}
$$
\caption{(Color online) The scaling plot of the correlation length 
at the critical point. Circles, squares, up-triangles, and down-triangles represent $T=2.281$ for $\alpha=0.0001$, 
$T=2.332$ for $\alpha=0.001$, $T=2.477$ for $\alpha=0.01$ and $T=2.984$ for $\alpha=0.1$, respectively. 
The linear system sizes are $L=20, 40, 80, 160,$ and $320$. 
The dashed line is proportional to $y=1/x$. 
The data are in excellent agreement with the scaling relation (\ref{scalingxialpha}).
}
\label{rik_scaling}
\end{figure}

Here we estimate the correlation length from the spin correlation function 
$c(r)=\langle\sigma_i\sigma_j\rangle$,
where $r$ is the distance between the sites $i$ and $j$, 
by the following definition
\beq
\xi(L)={\int_0^{L/2}\left(c(r)-c(L/2)\right)rdr\over \int_0^{L/2}\left(c(r)-c(L/2)\right)dr}.
\eeq 
 This definition gives the correlation length if $c(r)$ decays exponentially to its large-$r$ value, and also
for general cases 
it gives an estimate of the correlation length.
In Fig. \ref{Xi}, we depict a typical example of $c(r)$. At large distance, $c(r)$ is constant,\cite{Miya1} 
proportional to $\sqrt[]{N}$. The size dependence of $\xi(\alpha, L)$ is 
depicted in Fig. \ref{XialphaL}, where we confirm that the correlation length saturates for large $L$ as expected, 
even for quite small values of $\alpha$. 
The estimated $\xi(\alpha, L)$ 
are plotted in the finite-size scaling plot Fig. \ref{rik_scaling}, in which we assume 
$\xi$ at the critical point depends on $\alpha$ as (\ref{scalingxialpha}). 
We find that the data collapse onto a scaling function and thereby confirm the theoretical scaling relation (\ref{scalingxialpha}).

\section{Summary}
We found that in systems with both long and short-range interactions, 
the long-range interaction dominates the critical properties, even if it is 
infinitely weak.
At the critical temperature, although the susceptibility diverges, 
the cluster size does not. At the critical temperature, the system has 
a finite correlation length.

In this paper, we obtained a formula for the change of the critical 
temperature as a function of the strength $\alpha$ of the long-range interaction, 
and also a scaling form for the spin correlation length at the critical point.
 
The crossover of the nature of the order as the length scale changes was studied 
by a Monte Carlo method. We investigated the values of the Binder cumulant at its crossing points. 
It moved from the value of the short-range Ising model to that of the mean-field universality 
class, which enabled us to estimate the critical point systematically.  
The result agrees well with our proposed formula. 
We further note that our model can be considered as a gwell-stirredh approximation for the 
Ising model on a small-world network.\cite{Hast03}

We also proposed a scaling relation for the correlation length 
at the critical point as a function of $\alpha$.
At the critical point, the spin correlation function at large distances is constant,\cite{Miya1} 
proportional to $\sqrt[]{N}$ with a short-range component characteristic of the 
correlation length $\xi$. We obtained the value of the correlation length from the simulated spin correlation function, 
thus providing numerical confirmation of our proposed a scaling function.

We expect that the results found in this paper are applicable also 
for real system with degrees of freedom corresponding to lattice 
deformation. A study of such a model will be published elsewhere.\cite{Next}
We hope this kind of phenomena will be found in future experiments.

\section*{Acknowledgements}
The present work was supported by Grant-in-Aid for Scientific Research
on Priority Areas, and also the Next Generation Super Computer
Project, Nanoscience Program from MEXT of Japan.
The numerical calculations were supported by the supercomputer center of
ISSP of University of Tokyo. Work at Florida State University was supported in 
part by U.S. National Science Foundation Grant No. DMR-0802288.
T.M. acknowledges the support from JSPS (Grant No. 227835).

\appendix 
\section{The estimation of $T_c(\alpha)$ from the crossing points of the Binder cumulant}
\label{appendixA}
\begin{figure}[t]
$$
\includegraphics[scale=0.3]{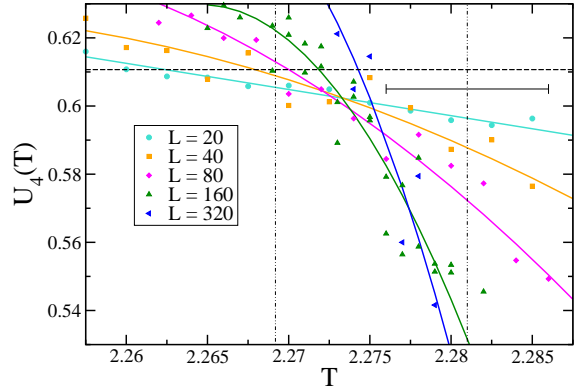} 
$$
\caption{
Detail of the crossings of the Binder cumulant for $\alpha=0.0001$. 
This figure is a magnified portion of Fig.~\protect\ref{Binderalpha}(d).}
\label{Binder} 
\end{figure}

In the present models, the system behaves like a short-range 
Ising model for small $\alpha$ and $L$.
If we study the crossing point of the Binder cumulants, $U_4(\alpha, L)$, for 
small values of $L$ and at small $\alpha$,
the crossing point gives a value close to that of the Ising model, i.e., 
$U_4^{*\rm IS}\simeq 0.61\dots$.
However, as the size increases, the crossing points approach the 
fixed-point value of the mean-field model,
$U_4^{*\rm HT}\simeq 0.27\dots$. 
For $\alpha=0.0001$, the crossing point of the two largest sizes 
simulated, $L=320$ and $160$, 
still stays near $0.57$, which is far from $U_4^{*\rm HT}$.
Thus, we cannot obtain the critical temperature directly.
Here we estimate $T_c(\alpha=0.0001)$ in the following way.
We obtain the crossing points for systems with $L$ and $L'=L/2$ 
as depicted in Fig.~\ref{Binder}.
Continuous lines for the cumulants as functions of $T$ were obtained as 
polynomial fits to densely spaced Monte Carlo data obtained from simulation 
runs of up to $10^7$~MCS. 
We assume the following properties:
(1) $U_4$ at the crossing point for large $L$ equals $U_4^{*\rm HT}$,
(2) $U_4(\alpha, L)$ is a monotonic function of the temperature, and
(3) the crossing temperature increases monotonically with $L$.
From these assumptions we find in Fig.~\ref{Binder} that 
$T_c(\alpha)$ is above $T=2.277$ which is the crossing temperature for 
$L'=160$ and $L=320$.
Because $U_4(\alpha, L=320)$ crosses $U_4^{*\rm HT}$ at $T=2.289$,
$T_c(\alpha)$ is below $T=2.289$.
By linear extrapolation with respect to $1/L$ 
of the crossing values for $L'/L = 80/160$ and $160/320$ (see the inset in Fig.~\ref{TCsize}), we estimated the 
critical temperature as
\beq
T_c(\alpha=0.0001)=2.281\pm 0.005.
\eeq


\end{document}